\documentclass[eprint]{actapoly}

\usepackage{multirow}

\newcommand{\apj}{ApJ}
\newcommand{\apjl}{ApJL}

\newcommand{\mnras}{MNRAS}

\newcommand{\nar}{NewAR}

\newcommand{\aap}{A\&A}

\begin{document}

\title[Galactic Center Minispiral:\\ Interaction Modes of Neutron Stars]
{Galactic Center Minispiral: Interaction Modes of Neutron Stars}

\correspondingauthor[Michal Zaja\v{c}ek]{Michal Zaja\v{c}ek}{ph1,max-planck,my,charles}{zajacek@ph1.uni-koeln.de}
\author[Vladim\'{i}r Karas]{Vladim\'{\i}r Karas}{my}
\author[Devaky Kunneriath]{Devaky Kunneriath}{my}

\institution{ph1}{I. Physikalisches Institut der Universit\"at zu K\"oln, Z\"ulpicher Strasse 77, D-50937 K\"oln, Germany} 
\institution{max-planck}{Max-Planck-Institut f\"ur Radioastronomie (MPIfR), Auf dem H\"ugel 69, D-53121 Bonn, Germany}
\institution{my}{Astronomical Institute, Academy of Sciences, Bo\v{c}n\'{\i}~II 1401, CZ-14131~Prague, Czech Republic}
\institution{charles}{Charles University in Prague, Faculty of Mathematics and Physics, V Hole\v{s}ovi\v{c}k\'ach 2, CZ-18000 Prague, Czech Republic}

\begin{abstract}
Streams of gas and dust in the inner parsec of the Galactic center form a distinct feature known as the Minispiral, which has been studied in radio waveband as well as in the infrared wavebands. A large fraction of the Minispiral gas is ionized by radiation of OB stars present in the Nuclear Star Cluster (NSC). Based on the inferred mass in the innermost parsec ($\sim 10^6$ solar masses), over $\sim 10^3$ -- $10^4$ neutron stars should move in the sphere of gravitational influence of the SMBH. We estimate that a fraction of them propagate through the denser, ionized medium concentrated mainly along the three arms of the Minispiral. Based on the properties of the gaseous medium, we discuss different interaction regimes of magnetised neutron stars passing through this region. Moreover, we sketch expected observational effects of these regimes. The simulation results may be applied to other galactic nuclei hosting NSC, where the expected distribution of the interaction regimes is different across different galaxy types.
\end{abstract}

\keywords{Galaxy: center, ISM: individual objects (Sagittarius~A), Stars: neutron}

\maketitle

\section{Introduction}

The Galactic center hosts the supermassive black hole (SMBH) observed as the compact radio source Sgr~A*, which is surrounded by the Nuclear star cluster (NSC) and gaseous-dusty structures, such as HII Minispiral arms of Sgr~A West, supernova remnant Sgr~A East, molecular clouds, and the Circumnuclear disk \citep{2010RvMP...82.3121G,2005bhcm.book.....E}. It is the closest SMBH and hence its environment can be studied with the highest resolution among galactic nuclei in the radio-, mm-, submm-, infrared, and X-ray wavebands \citep{2005bhcm.book.....E}. However, despite high-resolution multiwavelength studies several processes are still not satisfactorily explained, such as the star-formation near the SMBH, the feeding and feedback of Sgr~A*, and the distribution of the magnetic field and its interaction with other stellar and non-stellar components.  

The observations of the Galactic center region revealed a large population of young massive stars orbiting the SMBH as close as $\sim 0.1\,\rm{pc}$ \citep{2010ApJ...708..834B}. In fact, the NSC seems to be one of the densest concentrations of young massive stars in the Galaxy \citep{2010RvMP...82.3121G}. On the other hand, there is an observable flat distribution of late-type stars with a radius of as much as $10''$ \citep{2009A&A...499..483B,2009ApJ...703.1323D}. Thus, a steep relaxed Bahcall-Wolf cusp of stars with a slope of $7/4$ or $3/2$ \citep{1976ApJ...209..214B,1977ApJ...216..883B} is probably absent \citep{2009ApJ...703.1323D,2014A&A...566A..47S}.  

The estimates of the number of stellar remnants that use the power-law initial mass function (IMF) (standard Salpeter or top-heavy) combined with the mass segregation over the age of the bulge $(\sim 10\,\rm{Gyr})$ lead to a considerable population of stellar black holes of the order of $\sim 10^4$ \citep{1993ApJ...408..496M,2000ApJ...545..847M,2014MScT.........1Z}. The same order is expected for neutron stars based on multiwavelength statistical studies \citep{2012ApJ...753..108W}. Based on the total X-ray luminosity of the innermost parsec, \citet{2007MNRAS.377..897D} set an upper limit on the number of compact remnants ($\lesssim 40\,000$). 

Such an abundant population of neutron stars exhibiting strong magnetic fields could be utilized to further extend our knowledge about the processes in the Galactic center. The observations of neutron stars (pulsars as well as X-ray sources) near the SMBH would contribute to:
\begin{itemize}
\item our understanding of the star formation processes near the Galactic center using the number and the age distribution of observed sources,
\item mapping the gravitational potential near the SMBH using their period derivatives,
\item constraining the electron density profile in the Galactic center using their dispersion measures. 
\end{itemize}       

Despite continuing efforts only very few pulsars have been detected in the broader Galactic center region. It is thought that the lack of detections is due to profound interstellar dispersion and scattering. However, there are observational hints that such a population is present. \citet{2006MNRAS.373L...6J} report the discovery of two highly dispersed pulsars with the angular separation $\lesssim 0.3^{\circ}$ from the Galactic center. \citet{2009ApJ...702L.177D} confirm the detection of three pulsars with large dispersion measures with an offset of $\sim 10'$--$15'$ from Sgr~A*. There is an indication of the relation between the Sgr~A East SNR and the ``Cannonball'' source, which was detected in both the X-ray and radio wavebands and appears to be a pulsar wind nebula \citep[PWN;][]{2013ApJ...777..146Z}. This shows that neutron stars can form directly in the Galactic center environment. 

\begin{figure}
\centering
\includegraphics[width=\linewidth]{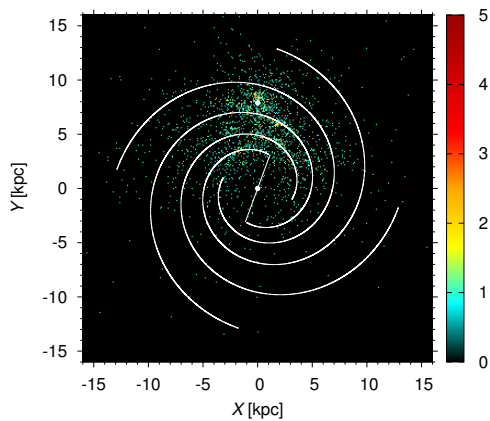} 
\caption{The distribution of 2302 pulsars in the $XY$ Galactic plane. Four logarithmic spiral arms are marked by white lines. Sgr~A* and the Sun are labelled as white points at position $(0,0)\,\rm{kpc}$ and $(0,7.9)\,\rm{kpc}$, respectively. The data are taken from the ATNF Pulsar Catalogue \url{http://www.atnf.csiro.au/research/pulsar/psrcat} \citep{2005AJ....129.1993M}.}
\label{fig:psrxy}
\end{figure}

In the innermost parsec only one magnetar PSR J1745-2900 was affirmed by independent detections \citep{2013ApJ...770L..24K,2013ApJ...770L..23M,2013ATel.5040....1E} at a projected distance of $\sim 2.4 \pm 0.3''$ from Sgr~A*, having a period of 3.7635537(2) \citep{2013ApJ...775L..34R}. The consequence of the lack of sensitivity to detect more neutron stars in the Bulge and the Galactic center is shown in Figure~\ref{fig:psrxy}, where most of the detected sources are concentrated relatively close to the Sun.  

In this contribution we aim to clarify the distribution of the interaction modes of neutron stars in the Galactic center. Based on the results we discuss the possibility of detecting neutron stars in the innermost parsec indirectly, specifically by looking for bow-shock structures similar to that of the ``Cannonball''  \citep{2013ApJ...777..146Z}.

The structure of the paper is as follows. In Section~\ref{sec:set-up} we explain the set-up of the model and the methods that are employed, including the introduction to a simple, analytical theory of interaction modes of neutron stars that is, however, sufficient for our purposes. Subsequently, in Section~\ref{sec:distribution} we study the distribution of interaction modes in the innermost parsec of the Galaxy. In Section~\ref{sec:discussion} we discuss the consequences of this distribution and the possibilities of detecting a fraction of the population indirectly. Finally, we summarize our conclusions in Section~\ref{sec:conclusions}.

\section{Set-up of the model and methods}
\label{sec:set-up}

In our model we concentrate on the innermost parsec of the Galactic center, which lies within the sphere of influence of the supermassive black hole (Sgr~A*) with the radius $r_{\rm{SI}}$:

\begin{equation}
r_{\rm{SI}}\approx 1.7 \left(\frac{M_{\bullet}}{4.0\times 10^6\,\rm{M_{\odot}}}\right)\left(\frac{\sigma}{100\,\rm{km\,s^{-1}}}\right)^{-2}\,,
\label{eq_sphereofinfluence}
\end{equation}
where $M_{\bullet}$ is the central black hole mass and $\sigma$ is the stellar velocity dispersion.

Since we are studying processes that occur at a distance of $\gtrsim 1000$ gravitational radii from the SMBH we approximate the gravitational field by a Newtonian point mass of $M_{\bullet}=4.0 \times 10^6\,M_{\odot}$ \citep{2010RvMP...82.3121G,2005bhcm.book.....E}. We employ a Monte Carlo approach for studying the rate of interactions between neutron stars and gaseous-dusty structures. Initially, we generate the orbits of neutron stars that do not gravitationally interact with each other and other stellar and non-stellar components of the NSC except for the SMBH. Hence, the orbital elements of individual stars do not change in the course of the simulation. Although this may seem an oversimplification, it is sufficient for statistical studies of the distribution of interaction modes.

\begin{figure}
\centering
\begin{tabular}{c}
\includegraphics[width=\linewidth]{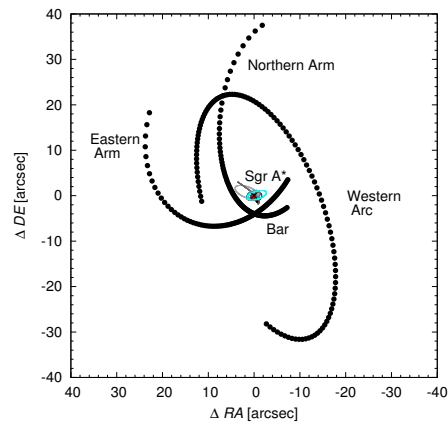}\\
\includegraphics[width=\linewidth]{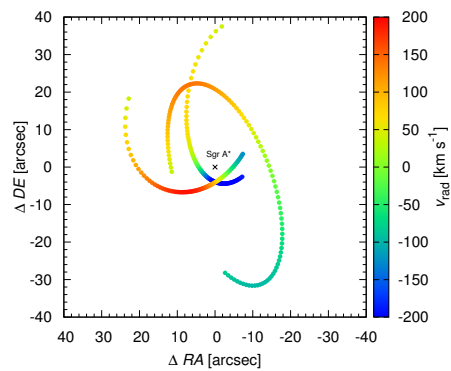}
\end{tabular}
\caption{The Minispiral model. Top panel: the nomenclature of the main components. Bottom panel: Keplerian velocity profile of the gaseous material according to \citep{2009ApJ...699..186Z,2010ApJ...723.1097Z}.}
\label{fig:minispiral}
\end{figure}

The thermal HII region of Sgr~A West known as the Minispiral has been studied in the radio-, mm-, and infrared wavelengths \citep{2012A&A...538A.127K,2009ApJ...699..186Z,2010ApJ...723.1097Z}. It consists of four main components: the Northern Arm, the Western Arc, the Eastern Arm, and the Bar \citep[see Figure~\ref{fig:minispiral}, ][]{2009ApJ...699..186Z,2010ApJ...723.1097Z,2012A&A...538A.127K}. The Minispiral consists of a mixture of ionised and neutral gas and dust, with temperature  ranging from $\sim 100\,\rm{K}$ for dust up to $10^4\,\rm{K}$ for the hot ionised gas phase. The inferred electron densities are $\sim 10^4$--$10^5\,\rm{cm^{-3}}$ \citep{2009ApJ...699..186Z,2010ApJ...723.1097Z}.

In our simplified model the Minispiral is represented as a system of spherical clumps along the three arms ($\sim 20$ for each arm) with the length-scale of each clump 1--2 arcsec. The density and temperature profiles are taken from \citep{2012A&A...538A.127K,2009ApJ...699..186Z,2010ApJ...723.1097Z}, but we also analyse the interaction modes for values outside this range. The velocity profile is assumed to be roughly Keplerian, consisting of three streams as inferred from observations by \citet{2009ApJ...699..186Z,2010ApJ...723.1097Z} (see Figure~\ref{fig:minispiral}). We note that deviations from Keplerian motion were also detected, probably due to magnetohydrodynamic effects \citep{2009ApJ...699..186Z,2010ApJ...723.1097Z}.   



\subsection{Interaction modes of magnetized neutron stars}
\label{subsec:interaction_modes}

Many observed neutron stars exhibit considerable magnetic fields, in the case of pulsars most frequently of the order of $10^{12}$ Gauss. The warm ionized gas that is also detected close to the Galactic centre, specifically in the Sgr~A West region \citep[e.g.][and references therein]{2012A&A...538A.127K} has high electric conductivity that is proportional to its temperature, $\lambda_{\rm{c}}\approx 10^7\, T_{\rm{e}}^{3/2}\,\rm{cm^{-1}}$. Therefore, this plasma must interact effectively with the large magnetic field of neutron stars. Consequently, we need magnetohydrodynamic (MHD) equations describing the gas dynamics in the potential of a neutron star. In fact, if we are close to the neutron star surface, we have to solve a system of relativistic magnetohydrodynamic equations (RMHD), which is often difficult for real systems.       

In our analysis, we focus on the fundamental characteristics of the interaction of a rotating magnetized neutron star with the plasma in the surroundings. This interaction consists of two parts: \textit{gravitational interaction} characterized by an accretion rate $\dot{M}$ of the captured medium; and \textit{electromagnetic interaction}, which is described by a magnetic dipolar moment $\mu$ and by a rotational period of a neutron star $P$. These three basic parameters, $\dot{M}$, $\mu$, and $P$, are further complemented by the mass of neutron star $M_{\rm{NS}}$ and the relative velocity with respect to the medium $v_{\infty}$. In fact, neutron stars are a part of the broader class of \textit{gravimagnetic rotators} characterized by mass $M$, angular momentum $\mathbf{J}=I\mathbf{\Omega}$, and the magnetic field, which is as the first approximation characterized by the dipole moment $\boldsymbol\mu$. 

In order to simplify the problem, we consider the following assumptions:
\begin{itemize}
 \item[(a)] the interaction takes place far from the neutron star surface, so a relativistic approach is not necessary,
 \item[(b)] the electromagnetic part of the interaction is independent of the accretion flux parameters,
 \item[(c)] the intrinsic magnetic field of neutron stars is a dipole field.
\end{itemize} 

\begin{figure}[h!]
\centering
\includegraphics[width=\linewidth]{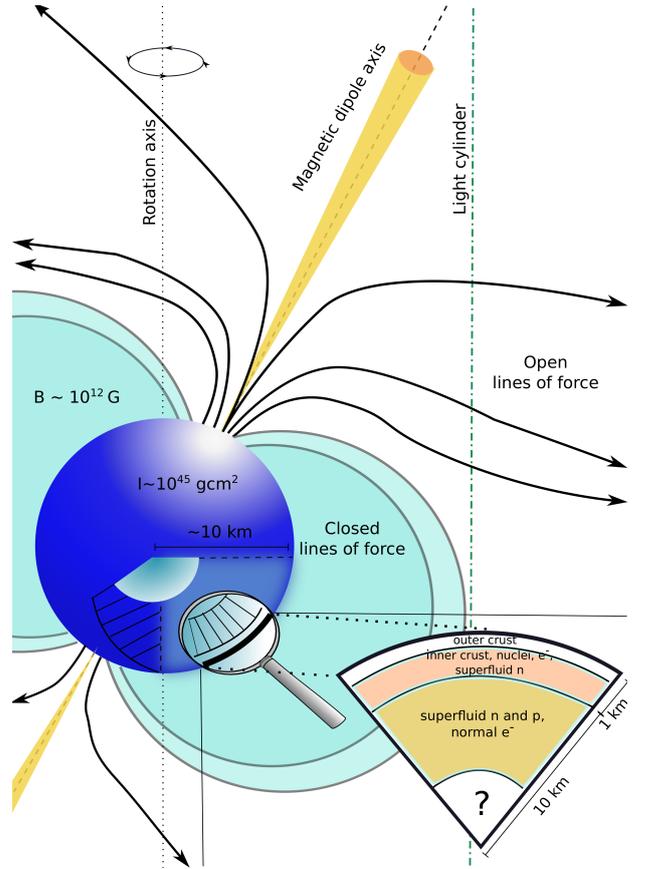}
\caption{Sketch of a canonical pulsar. In the lower right corner, a cross-section of the neutron star is depicted with two basic components: the crust and the superfluid neutron interior. Magnetic field lines are closed up to the light cylinder. Closer to the magnetic pole, field lines open and relativistic particles escape.}
\label{img_neutron_star_canonical}
\end{figure}

 Let us make a few notes concerning assumptions (a), (b), and (c). Point (a) means that the interaction of interstellar plasma with the magnetic field or a relativistic particle wind of the neutron star takes place at $r \gg GM_{\rm{NS}}/c^2$, which is often the case. Assumption (b) means that an infalling plasma does not distort the intrinsic magnetic field of the neutron star significantly. Point (c) is approximately valid due to the conservation of the magnetic flux. The ratio of the quadrupole $q$ and dipole $\mu$ components of the field scales with the stellar radius as $q/\mu \propto R$ during the core-collapse, so the field is effectively cleansed of higher multipole components. The dipole field is the most representative component far away from the surface, where the interaction with the ambient medium takes place. The closer to the surface the matter gets, the more important the quadrupole and the higher multipole components become.
 
 Given assumptions (a), (b), (c), the problem becomes much simpler and many effects of the interaction are thus neglected. However, one still gets basic information about the character and the scale of the interaction if fundamental parameters of the neutron star and those of the ambient medium are given.

For a non-rotating neutron star, the harmonic surface current $j\approx \sin{\theta}$ induces a dipole field outside the sphere having the following form:

\begin{equation}
\mathbf{B_{\rm{d}}}=\frac{2\mu \sin{\theta}}{r^3}\mathbf{e_{\rm{r}}}-\frac{\mu \cos{\theta}}{r^3}\mathbf{e_{\rm{\theta}}}\,,
\label{eq_magfield_vector}
\end{equation} 
where $\mathbf{e_{\rm{r}}}$ and $\mathbf{e_{\rm{\theta}}}$ are unit vectors, $\mu$ denotes the dipole moment. The angle $\theta$ is measured from the axis perpendicular to the dipole field axis (see the sketch in Figure~\ref{img_neutron_star_canonical}). The magnetic field magnitude $B_{0}$ at the poles is twice as big as at the equator, which follows from the magnitude relation $B_{\rm{d}}=\mu/r^3(1+3\sin^2{\theta})^{1/2}$. The magnetic dipole moment is given by the field magnitude at the poles $B_{0}$ and the neutron star radius $R_{\rm{NS}}$, $\mu=1/2B_0 R_{\rm{NS}}^3$, yielding typical values $\mu \approx 1/2 (2\times10^{12}\ \rm{Oe})\times 10^{18}\,\rm{cm^3} = 10^{30}\,\rm{Oe\,cm^{3}}$.\footnote{1 Oe is a unit of the $\mathbf{H}$ field in cgs units. The equivalent in SI units is 1 $\rm{A/m}$, $1\,\rm{A/m}=(4\pi\, 10^{-3})\,\rm{Oe}$.} 

In the case of a rotating dipole, the electric component of the electromagnetic field also has to be considered. If the rotational axis of a neutron star is inclined with respect to the dipole axis by angle $\alpha$, the neutron star emits electromagnetic dipole radiation at the rotational frequency $\Omega$. In the frame of this model, there exists the distance $R_{\rm{l}}=c/\Omega$, below which the electromagnetic field is static and the electric component is given by $E\approx v/c\,B=(\Omega r/c)B = (r/R_{\rm{l}})B$. While approaching the radius of the light cylinder, $R_{\rm{l}}$, the electric component becomes comparable to the magnetic component. On the crossing of the light cylinder, the electromagnetic field is no longer static and it becomes a propagating electromagnetic wave. The magnetic dipole radiation carries away energy that may be written in terms of the magnetic dipole $\mu$,

\begin{equation}
\frac{\mathrm{d}E_{\rm{rad}}}{\mathrm{d}t}=\frac{2}{3c^3}\mu^2 \Omega^4 \sin^2{\alpha}\,.
\label{eq_magdipole} 
\end{equation}

\paragraph{Gaseous environment and possible modes of accretion.} A neutron star is assumed to pass through an ideally conducting plasma of density $\rho_{\infty}$, temperature $T_{\infty}$, and the sound speed $c_{\infty}$ at infinity. We also consider the relative motion of a star with respect to the surrounding medium, $v=v_{\star}$. The plasma starts falling onto the neutron star due to its attraction. In the case of neutron stars with a negligible magnetic field, the stationary flow of matter, or the capture rate, is given by the Bondi-Hoyle-Lyttleton relation \citep[e.g.,][]{zelʹdovich1971relativistic}:

\begin{equation}
\dot{M}_{\rm{c}}=\delta\frac{(2GM_{\rm{NS}})^2}{(v_{\star}^2+c_{\infty}^2)^{3/2}}\rho_{\infty}\,,
\label{eq_bhl2}
\end{equation} 
where $\delta$ is a dimensionless factor of the order of unity. Defining the capture cross-section $\sigma_{\rm{G}}=\delta \pi R_{\rm{G}}^2$, with $R_{\rm{G}}$ denoting the capture radius,

\begin{equation}
 R_{\rm{G}}=\frac{2GM_{\rm{NS}}}{v_{\star}^2+c_{\infty}^2}\,,
\label{eq_capture_radius}
\end{equation} 
  we may use a convenient form of the capture rate:

\begin{equation}
\dot{M}_{\rm{c}}=\sigma_{\rm{G}}\rho_{\infty}v_{\star}\,.
\label{eq_capture_rate}
\end{equation}

In realistic astrophysical problems, quantities $T_{\infty}$, $\rho_{\infty}$, and $c_{\infty}$ are taken at finite distances, $R \gg R_{\rm{G}}$.

Possible accretion modes may be divided into three distinct groups:

\begin{itemize}
\item  $v_{\star} \ll c_{\infty}$; spherical accretion, without angular momentum in accreting matter,
\item  $v_{\star} \gg c_{\infty}$; cylindrical accretion -- axially-symmetric accretion, without angular momentum in the accreting matter,
\item the disk accretion, infalling matter has a considerable angular momentum.
\end{itemize}

The following critical radii can be distinguished in agreement with the terminology of \citep{lipunov1992}.

\paragraph*{Stopping radius.}

Neutron stars as gravimagnetic rotators attract ionized matter due to gravitational forces on the one hand, and they prevent accretion due to electromagnetic forces on the other. In the rotating dipole model, the electromagnetic field is stationary inside the light cylinder, $R_{\rm{l}}=c/\Omega$, and changes into a freely propagating electromagnetic wave beyond it.

 The luminosity $L_{\rm{m}}$ of the magnetic dipole radiation may be estimated by eq. \eqref{eq_magdipole}, $L_{\rm{m}}=\mathrm{d}E_{\rm{rad}}/\mathrm{d}t$.   \citet{1969ApJ...157..869G} found out that near the magnetic axis of the neutron star, which is inclined by a small angle with respect to the rotational axis, the electric component is directed along the magnetic field, $E\approx (\Omega r/c)B_0$, and it accelerates charged particles beyond the light cylinder up to relativistic energies, which effectively forms a \textit{pulsar wind}. It is assumed that this wind becomes frozen in the surrounding medium and passes its impulse to it. 
 
 The pressure of the wind may be estimated as $P_{\rm{ej}}=L_{\rm{m}}/(4\pi r^2 v_{\rm{ej}})$. However, it may  happen that the accreted plasma penetrates into the region surrounded by the light cylinder. It is prevented from accretion onto the surface by the pressure of the static magnetic field, $P_{\rm{m}}=B^2/(8\pi)=\mu^2/(8\pi r^6)$. To sum up, the pressure of electromagnetic forces is of a different character inside and outside the light cylinder, and may be approximated by the following relations: 

\begin{equation}
P_{\rm{m}} =
\begin{cases}
\frac{\mu^2}{8\pi r^6} & \text{if } r \leq R_{\rm{l}} \\
\frac{L_{\rm{m}}}{4 \pi r^2 c} & \text{if } r > R_{\rm{l}}\,.
\end{cases}
\label{eq_elmag_pressure}
\end{equation}

Using the light cylinder radius, $R_{\rm{l}}$, we may rewrite the dipole radiation luminosity of the neutron star $L_{\rm{m}}$, eq. \ref{eq_magdipole}, into the form:

\begin{equation}
L_{\rm{m}}=\kappa_{\rm{t}}\frac{\mu^2}{R_{\rm{l}}^3}\Omega\,,
\label{eq_magdipole2}
\end{equation}  
where $\kappa_{\rm{t}}=2/3\sin^2{\alpha}$ is a dimensionless factor. Using eq. \eqref{eq_magdipole2}, the electromagnetic pressure acting on the surrounding gas expressed by \eqref{eq_elmag_pressure} may be rewritten into the following relations:

\begin{equation}
P_{\rm{m}} =
\begin{cases}
\frac{\mu^2}{8\pi r^6} & \text{if } r \leq R_{\rm{l}} \\
\frac{\kappa_{\rm{t}}\mu^2}{4 \pi R_{\rm{l}}^4 r^2 } & \text{if } r > R_{\rm{l}}\,.
\end{cases}
\label{eq_elmag_pressure2}
\end{equation}
Electromagnetic pressure $P_{\rm{m}}$ is continuous for $\kappa_{\rm{t}}=1/2$ at $r=R_{\rm{l}}$.

Matter being accreted onto the neutron star exerts pressure, which is approximately constant beyond the gravitational capture radius $R_{\rm{G}}$ and is equal to $1/2 \rho_{\infty} v_{\star}^2$. For radii smaller than capture radius $R_{\rm{G}}$, matter falls almost freely and exerts dynamic pressure $1/2 \rho(r) v(r)^2$. Under the assumption of spherical accretion, $\dot{M}_{\rm{c}}=4 \pi r^2 \rho(r) v(r)$, which becomes $\dot{M}_{\rm{c}}=4 \pi R_{\rm{G}}^2 \rho_{\infty} v_{\star}$ beyond the capture radius, the accretion pressure $P_{\rm{a}}$ can be estimated as follows:

\begin{equation}
P_{\rm{a}} =
\begin{cases}
\frac{\dot{M}_{\rm{c}} v_{\star}}{8 \pi R_{\rm{G}}^2} & \text{if } r > R_{\rm{G}} \\
\frac{\dot{M}_{\rm{c}}v_{\star}}{8 \pi r^2}\left(\frac{R_{\rm{G}}}{r}\right)^{1/2} & \text{if } r \leq R_{\rm{G}}\,.
\end{cases}
\label{eq_acc_pressure}
\end{equation}
The accretion pressure is a continuous function according to eq. \eqref{eq_acc_pressure}.

\begin{figure*}[tbh!]
\centering
\begin{tabular}{cc}
\includegraphics[width=0.45\linewidth]{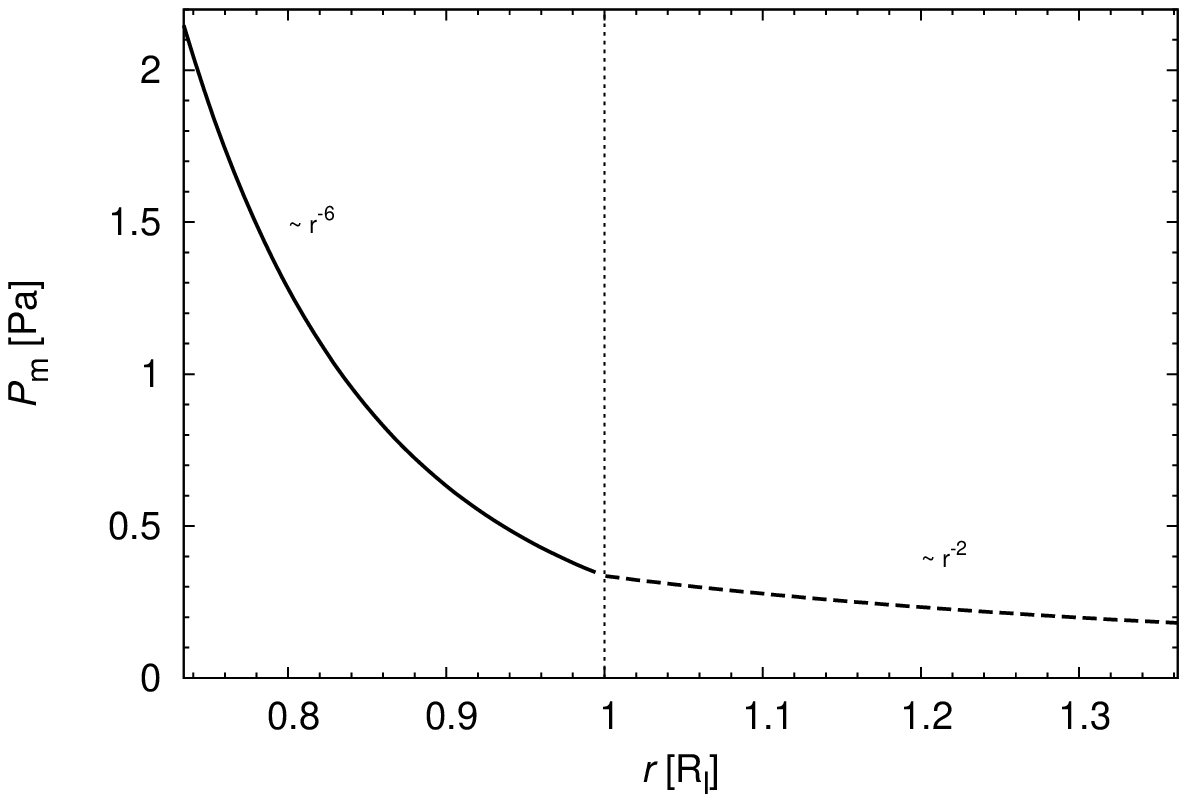} & \includegraphics[width=0.45\linewidth]{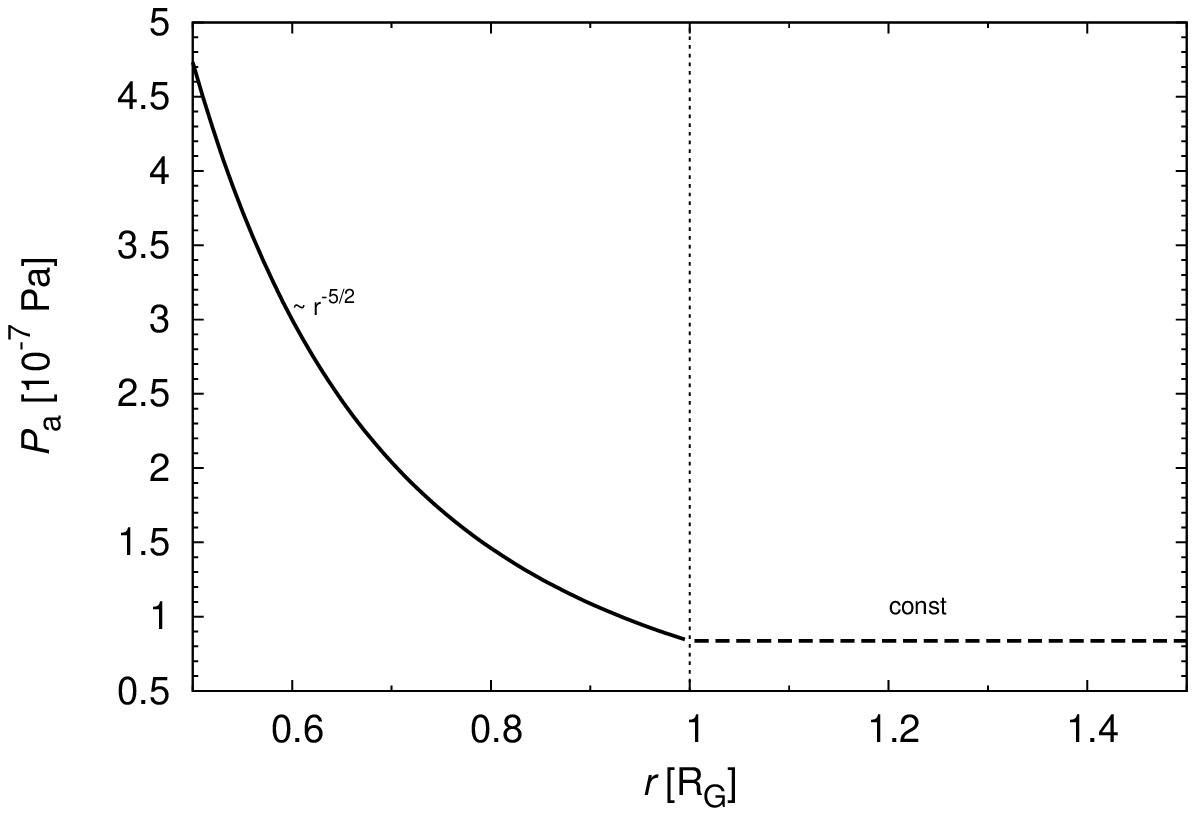}
\end{tabular}
\caption{Left panel: the pressure exerted by a stationary magnetic field varies as $\sim r^{-6}$, whereas outside the light cylinder the pressure of relativistic particles that act on infalling plasma falls off with distance as $\sim r^{-2}$. The distance is expressed in light radii ($R_{\rm{l}}$). Right panel: the pressure caused by accreted plasma is approximately constant beyond the gravitational capture radius, whereas for smaller radii gas falls freely and the pressure increases as $\sim r^{-5/2}$ while approaching the star. The units are expressed in gravitational capture radii ($R_{\rm{G}}$).}
\label{img_acc_elmag_pressure}
\end{figure*}

Both the electromagnetic pressure $P_{\rm{m}}$ and the accretion pressure $P_{\rm{a}}$ as functions of distance are plotted in Figure \ref{img_acc_elmag_pressure} for the following set of parameters: $P=1\,\rm{s}$, $B_0=2\times 10^{12}\,\rm{G}$, $R_{0}=10^6\,\rm{cm}$, $n_{\rm{H}}=10^4\,\rm{cm^{-3}}$, $T_{\rm{e}}=6\times     
 10^3\,\rm{K}$, $M_{\rm{NS}}=1.4\,M_{\odot}$, $v_{\star}=100\,\rm{km\,s^{-1}}$.  

The plasma being accreted by a neutron star is halted at the \textit{stopping radius} where the pressure of the electromagnetic forces is in balance with the pressure of the accreted matter. The stopping radius may be derived from the following condition:

\begin{equation}
P_{\rm{m}}=P_{\rm{a}}\,.
\label{eq_equality_elmag_accretion}
\end{equation}

If the pressure of electromagnetic forces is represented by the pressure of a static dipole field, then eq. \eqref{eq_equality_elmag_accretion} yields the Alfv\'{e}n radius, $R_{\rm{A}}$. If the electromagnetic pressure $P_{\rm{m}}$ is related to the outflow of relativistic particles (pulsar wind), then the corresponding radius is the so-called Shvartsman radius, $R_{\rm{Sh}}$. The stopping radius may therefore be expressed as \citep{lipunov1992}:

\begin{equation}
R_{\rm{st}} =
\begin{cases}
 R_{\rm{A}} & \text{if } R_{\rm{st}} \leq R_{\rm{l}} \\
 R_{\rm{Sh}} & \text{if } R_{\rm{st}} > R_{\rm{l}}\,.
\end{cases}
\label{eq_stopping_radius}
\end{equation} 

For the case when $R_{\rm{st}} \leq R_{\rm{l}}$, the Alfv\'{e}n radius is given by the corresponding relations in eqs. \eqref{eq_elmag_pressure2} and \eqref{eq_acc_pressure}, so again we get two cases for distances larger or smaller than the capture radius $R_{\rm{G}}$:

\begin{equation}
R_{\rm{A}} =
\begin{cases}
\left(\frac{4 \mu^2 G^2 M_{\rm{NS}}^2}{\dot{M}_{\rm{c}} v_{\star}^5}\right)^{1/6}  & \text{if } R_{\rm{A}} >  R_{\rm{G}} \\
\left(\frac{\mu^2}{\dot{M}_{\rm{c}}(2GM_{\rm{NS}})^{1/2}}\right)^{2/7}  & \text{if } R_{\rm{A}} \leq R_{\rm{G}}\,.
\end{cases}
\label{eq_alfven_radius}
\end{equation} 

For $R_{\rm{st}} > R_{\rm{l}}$, the relativistic pulsar wind interacts with the accreted matter. By comparing the dependencies in Figure \ref{img_acc_elmag_pressure}, we see that the accretion pressure increases as $\sim r^{-5/2}$ for $R_{\rm{st}} \leq R_{\rm{G}}$ and thus more rapidly than the pressure of the pulsar wind, which decreases as $\sim r^{-2}$. Hence, for $R_{\rm{st}} \leq R_{\rm{G}}$, no stable cavern can be maintained by the ejection of matter. However, using eq. \eqref{eq_elmag_pressure2} for $r>R_{\rm{l}}$ and eq. \eqref{eq_acc_pressure} for $r>R_{\rm{G}}$, we may find the stopping radius for distances $R_{\rm{st}} > R_{\rm{G}}$. This radius is also known as the Shvartsman radius and may be expressed in the following way:

\begin{align}
R_{\rm{Sh}} & = \left(\frac{2 L_{\rm{ej}}}{\dot{M}_{\rm{c}} v_{\star} v_{\rm{ej}}} \right)^{1/2} R_{\rm{G}}\,,\notag\\
R_{\rm{Sh}} & = \left(\frac{8 \kappa_{\rm{t}} \mu^2 (GM_{\rm{NS}})^2 \Omega^4}{\dot{M}_{\rm{c}}v_{\star}^5 c^4}\right)^{1/2}\,\text{if}\hspace{0.5cm} v_{\rm{ej}}=c\,.
\label{eq_shvartsman_radius}
\end{align}  

Let us note that the relations \eqref{eq_alfven_radius} and \eqref{eq_shvartsman_radius} are valid for accretion rates below the Eddington limit for accretion, $G\dot{M}_{\rm{c}}M_{\rm{NS}}/R_{\rm{st}}<L_{\rm{Edd}}$. The supercritical regime occurs in disk accretion modes that do not develop when neutron stars are just passing through an ionized plasma medium that has low angular momentum.

\paragraph*{Corotation radius.} The corotation radius of a rotating neutron star is another important distance scale. If accreting plasma penetrates beyond the light cylinder, it is stopped at the Alfv\'{e}n radius, $R_{\rm{st}}\approx R_{\rm{A}}$, where the plasma pressure and the pressure of the static magnetic field are in balance. Further evolution of stopped plasma is given by the rotational velocity of the neutron star. Let us assume that the plasma becomes frozen at $R_{\rm{st}}$ in  the magnetic field and corotates with it at the angular velocity of the neutron star, $\Omega$. The plasma clump  will eventually reach the surface of the neutron star if the rotational velocity at the stopping radius $R_{\rm{st}}$ is smaller than the Keplerian velocity at the same distance:

\begin{equation}
\Omega R_{\rm{st}} < \left(\frac{GM_{\rm{NS}}}{R_{\rm{st}}}\right)^{1/2}\,.
\label{eq_conditionforfalling}
\end{equation}

However, if the condition \eqref{eq_conditionforfalling} is not met, a centrifugal barrier develops that prevents plasma from accreting further. The critical corotation radius $R_{\rm{c}}$ that separates the two modes is given by the equality in \eqref{eq_conditionforfalling}:

\begin{equation}
R_{\rm{c}}=\left(\frac{GM_{\rm{NS}}}{\Omega^2}\right)^{1/3}\,.
\label{eq_corotationradius}
\end{equation}

If $R_{\rm{st}}<R_{\rm{c}}$, rotation does not considerably effect the accretion flux. If $R_{\rm{st}} \ge R_{\rm{c}}$, stationary accretion is not allowed. 

\begin{table*}[tbh!]
\centering
\resizebox{\textwidth}{!}{
\begin{tabular}{cccc}
\hline
\hline
Name & Notation & Relation between distances & Observational effects\\
\hline
Ejector & E & $R_{\rm{st}}>\rm{max}\{R_{\rm{G}},R_{\rm{l}}\}$ & radiopulsars \\
Propeller & P & $R_{\rm{c}}<R_{\rm{st}} \leq \rm{max}\{R_{\rm{G}},R_{\rm{l}}\}$ & spinning-down more efficient, transient sources\\
Accretor & A & $R_{\rm{st}} \leq R_{\rm{G}}$ and $R_{\rm{st}} \leq R_{\rm{c}}$ & X-ray pulsars, X-ray bursters \\ 
\hline
\end{tabular}
}
\caption{Summary of the interaction modes and the types of neutron stars that appear to be relevant in different regimes within the environment of the Minispiral.}
\label{tab_interaction_modes}
\end{table*}

\begin{figure*}[tbh!]
\centering
\begin{tabular}{ccc}
\includegraphics[width=0.3\linewidth]{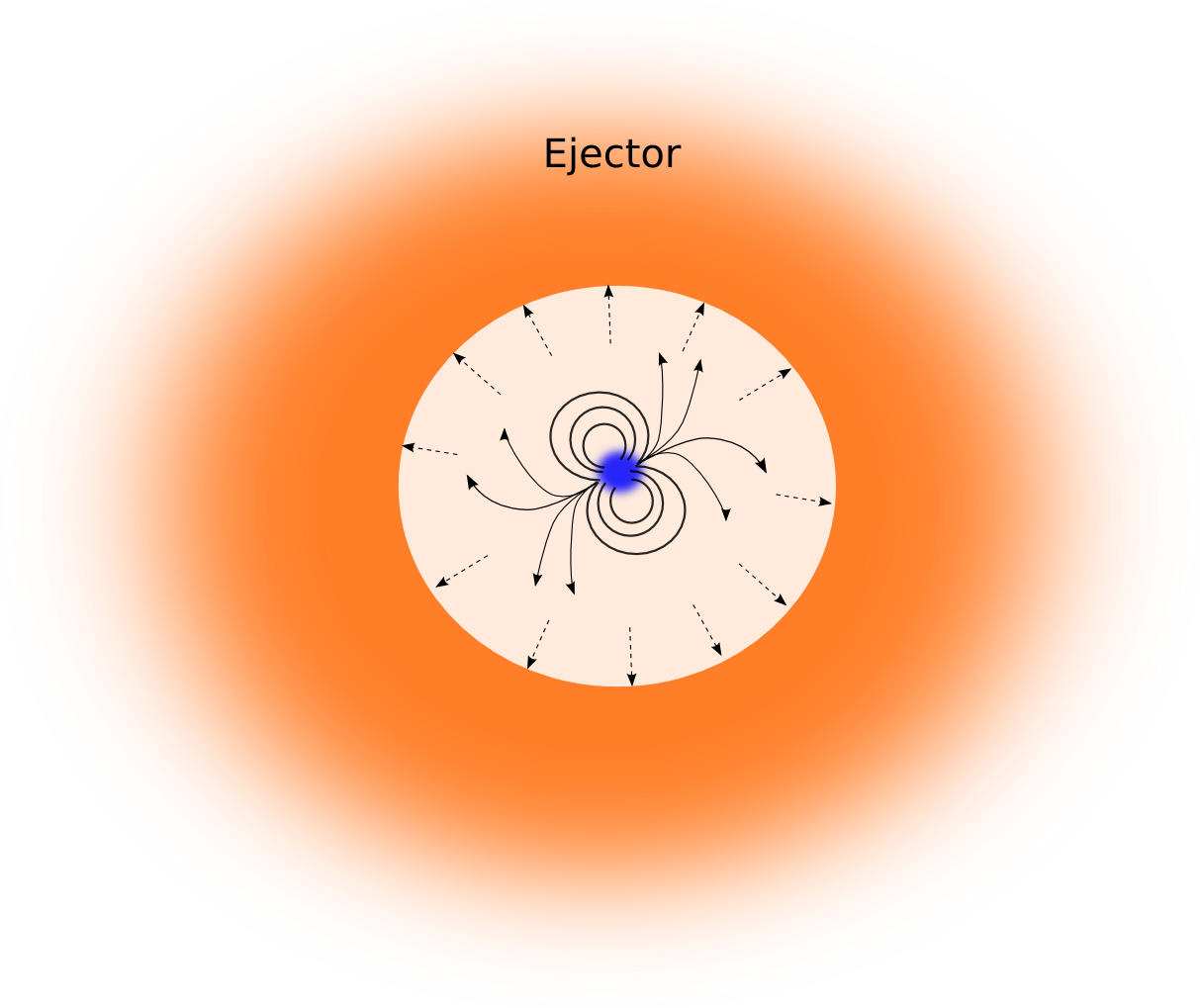} & \includegraphics[width=0.3\linewidth]{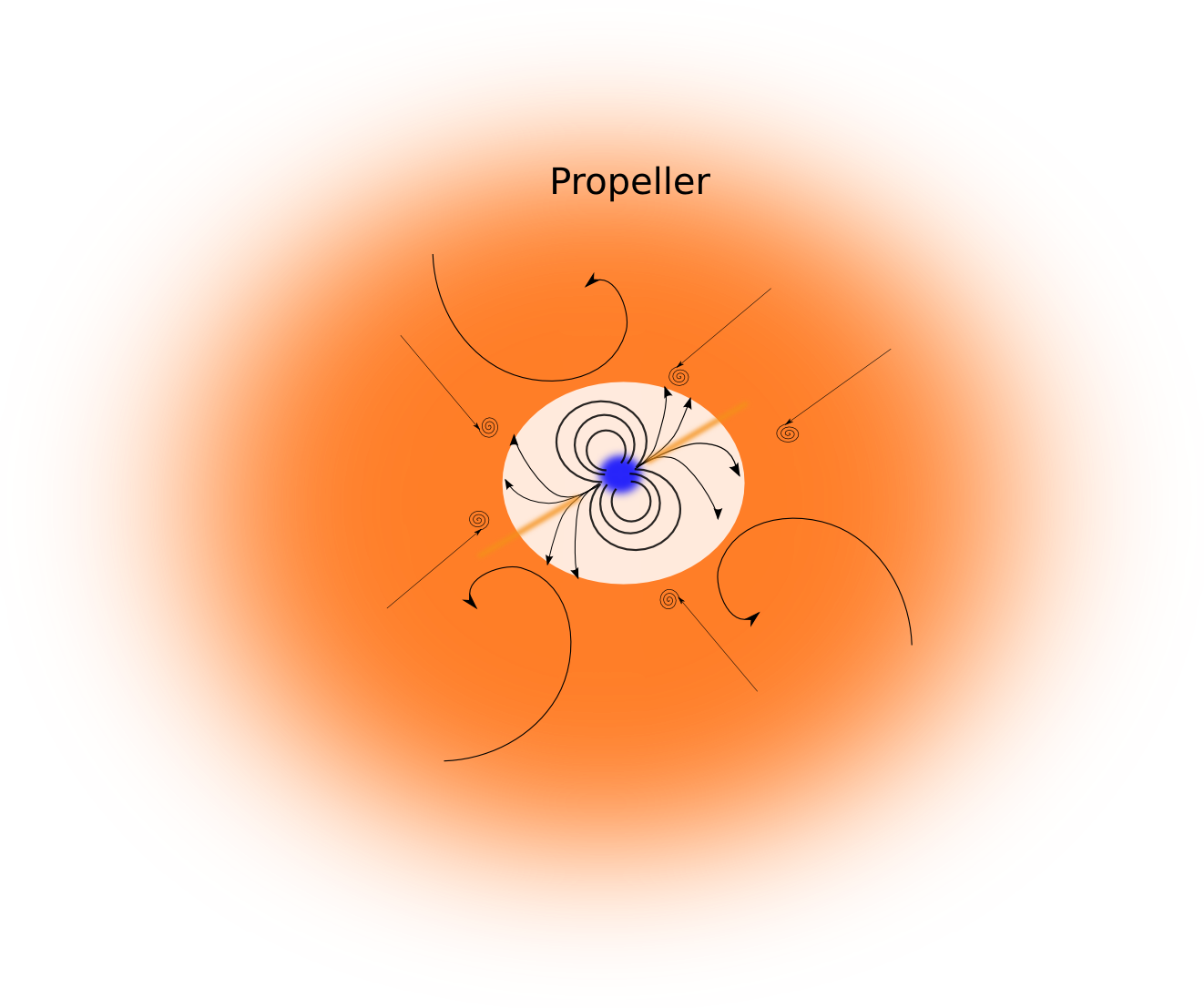} & \includegraphics[width=0.3\linewidth]{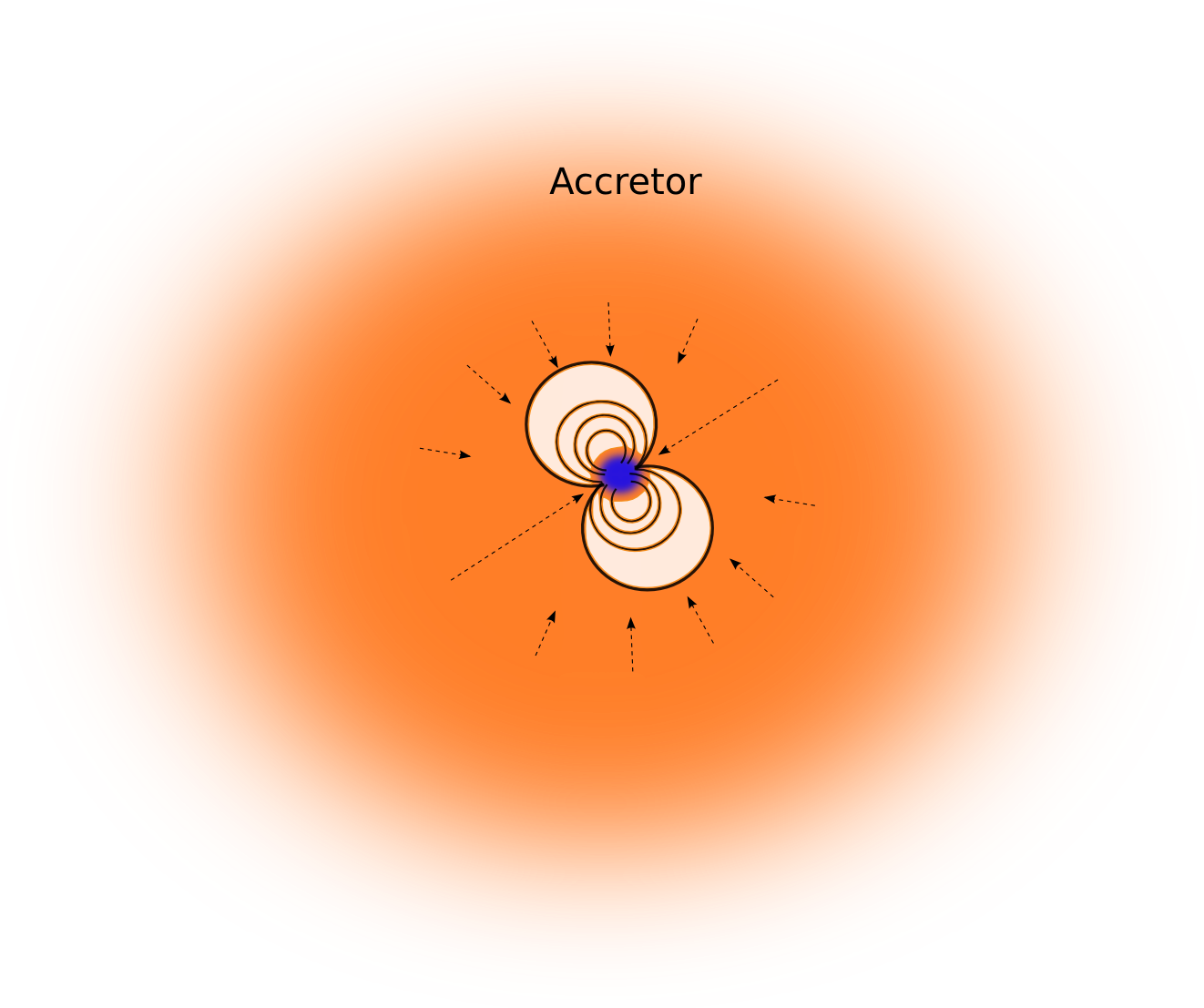} 
\end{tabular}
\caption{Illustrations of basic interaction modes of magnetized neutron stars with the surrounding environment: ejector, propeller, accretor.}
\label{img_intmodes}
\end{figure*}

\paragraph{Interaction modes.} The mode of interaction of magnetized rotating neutron stars and also other gravimagnetic rotators is given by the relations  between four fundamental distance scales: stopping radius $R_{\rm{st}}$, light cylinder radius $R_{\rm{l}}$, gravitational capture radius $R_{\rm{G}}$, and corotation radius $R_{\rm{c}}$. Elementary combinatorics yields $4!=24$ possible relations between these radii. However, when taking astrophysics into consideration, we get just a few modes. For example, $R_{\rm{l}}$ is always greater than $R_{\rm{c}}$. We do not take supercritical regimes into account in our analysis, since they occur in systems with disk accretion, and binary systems are also not considered. The classification of three main interaction modes -- ejector (E), propeller (P), and accretor (A) -- and thus types of neutron stars according to \citet{lipunov1992} is summarized in Table \ref{tab_interaction_modes}. The sketches of these basic regimes are shown in Figure \ref{img_intmodes}.

\section{Results: Distribution of interaction modes in the Galactic center}
\label{sec:distribution}

In our Monte Carlo simulations we produce a sufficiently large synthetic population of neutron stars. The number of members of this population is of the order of $10^4$--$10^6$. As a nominal distribution of period $\log{P}$ and  surface dipole magnetic field $\log{B}$ we take the Gaussian fit of the main peak of the observed distributions in the catalogue by  \citet{2005AJ....129.1993M}

\begin{align}
  N(\log{P}) \propto \exp{\left(-\frac{(\log{P}-\mu_{P})^2}{2\sigma_{P}^2}\right)}\,,\notag\\
  N(\log{B}) \propto \exp{\left(-\frac{(\log{B}-\mu_{B})^2}{2\sigma_{B}^2}\right)}\,, 
\label{eq_fits}    
\end{align}
where $\mu_P$ and $\mu_B$ are mean values of the Gaussian distribution and $\sigma_P^2$ and $\sigma_B^2$ are the corresponding variances.
The values of the period and surface magnetic field are independently attributed to each neutron star according to the distribution with the parameters in Table~\ref{table_dist1}.

\begin{table}[tbh!]
\centering
\begin{tabular}{ccc}
\hline
\hline
Quantity & $\mu$ & $\sigma$ \\
\hline
$\log{P}$ & -0.2188 & 0.3488\\
$\log{B}$ & 12.0900 & 0.4711\\
\hline 
\end{tabular}
\caption{Parameters of the synthetic distribution of the period (in seconds) and the synthetic distribution of the surface magnetic field (in Gauss).}
\label{table_dist1}
\end{table}

First, we fix the mean temperature of the ionized gas, $\overline{T}_{\rm{e}}= 6 \times 10^3\,\rm{K}$. The density is continually increased by an order of magnitude, starting with $10^4\,\rm{cm^{-3}}$; we consider densities up to $10^9\,\rm{cm^{-3}}$. 

\begin{figure*}[tbh!]
\centering
\begin{tabular}{cc}
\includegraphics[width=0.5\linewidth]{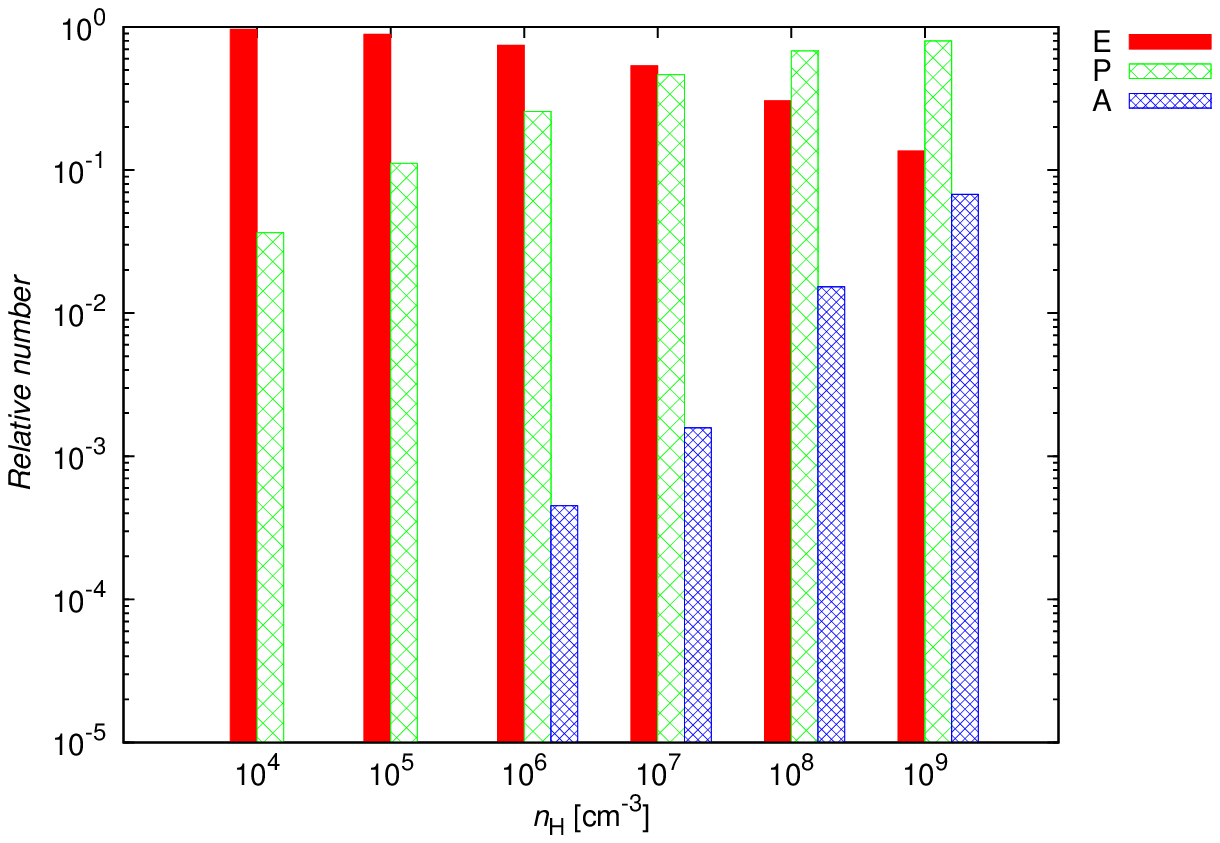} & \includegraphics[width=0.5\linewidth]{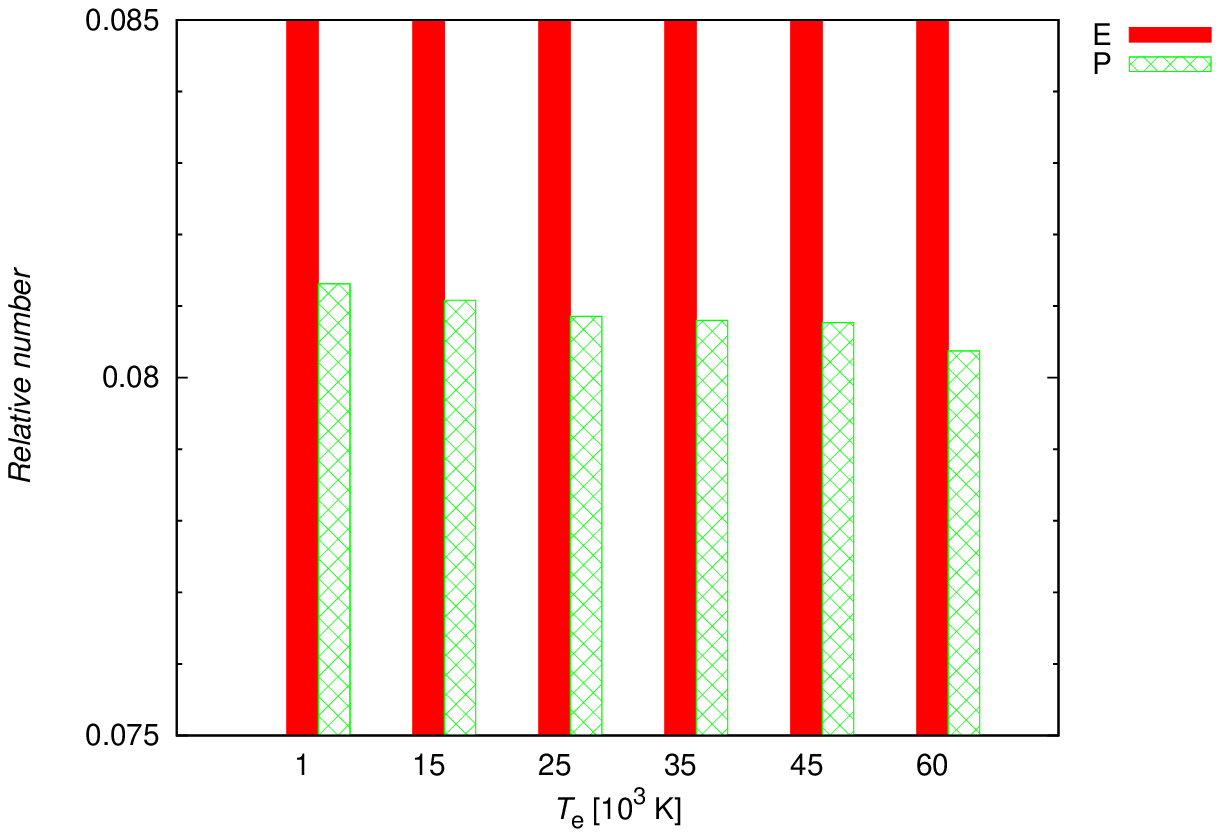}\\
\includegraphics[width=0.5\linewidth]{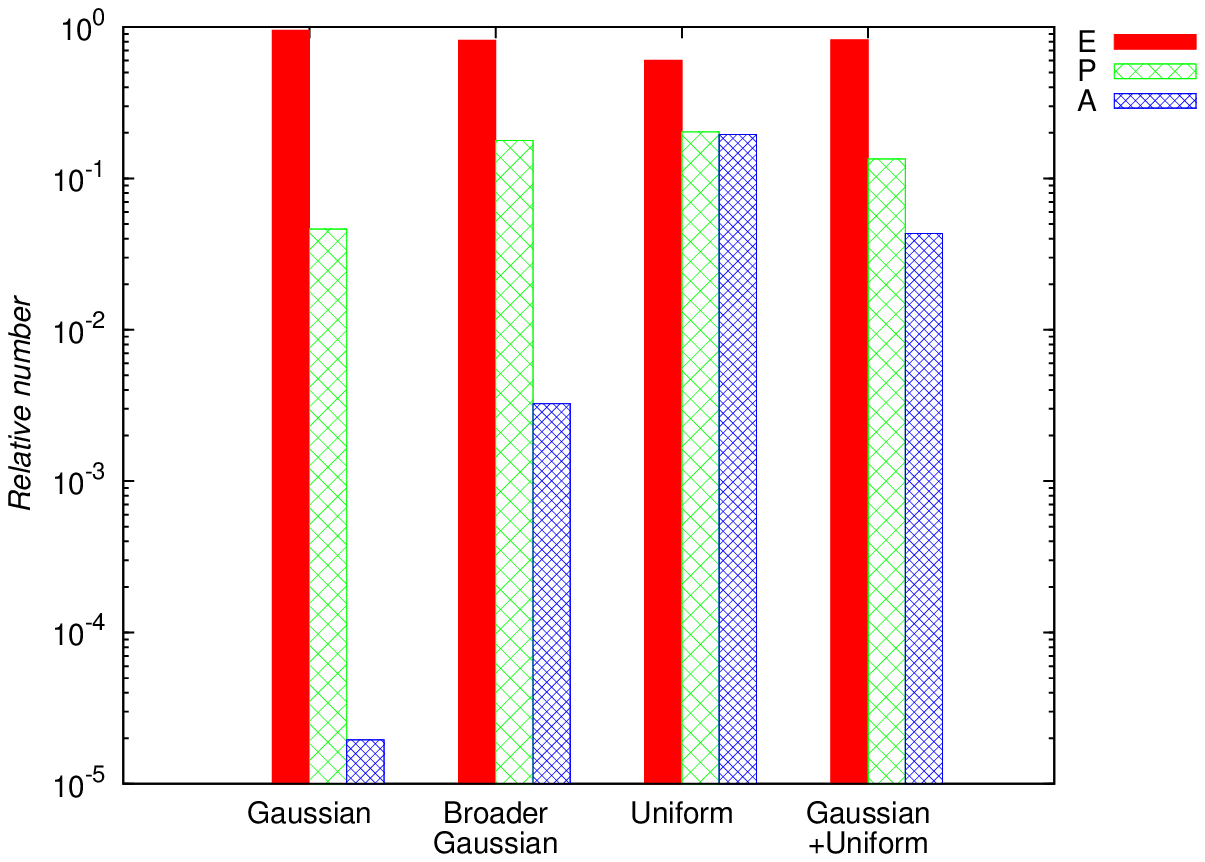} & \includegraphics[width=0.5\linewidth]{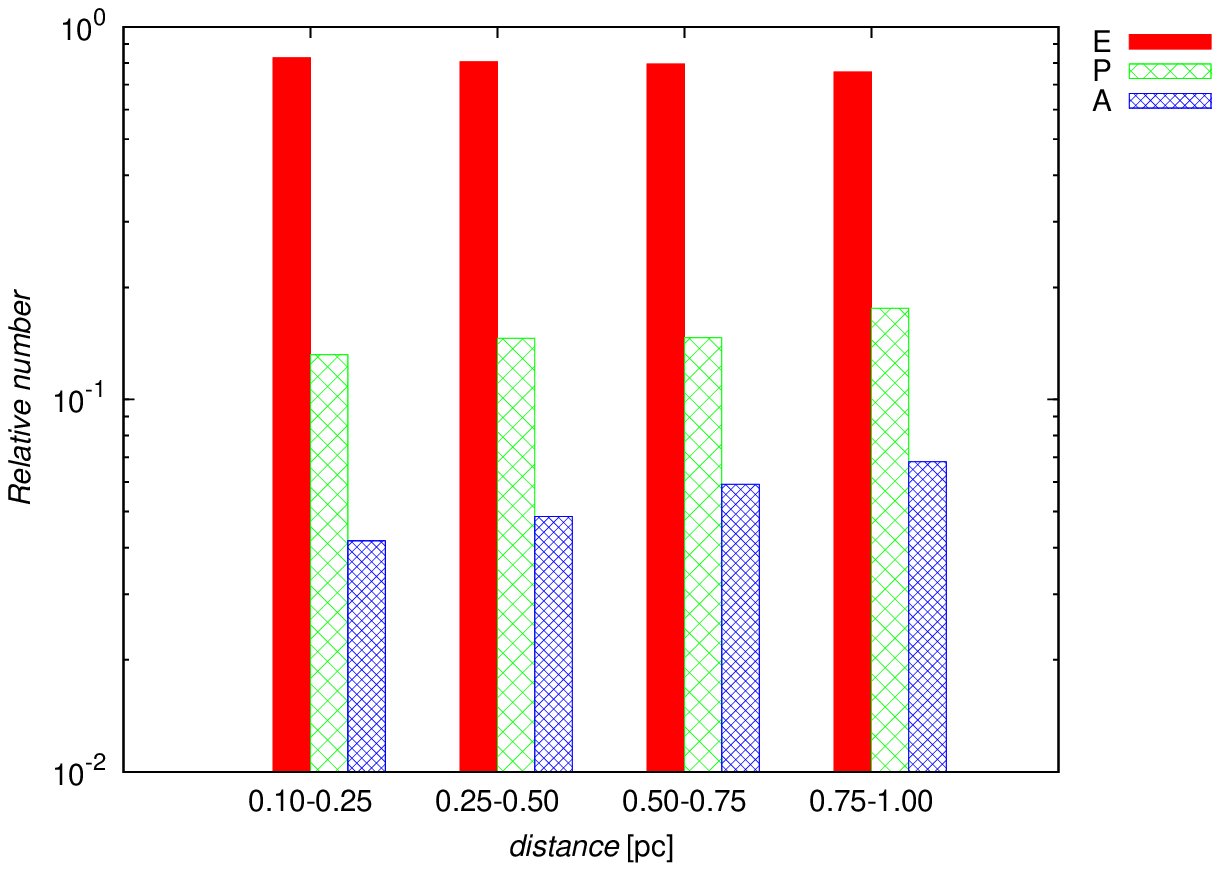}
\end{tabular}
\caption{Histograms of the relative distribution of interaction modes of neutron stars in the Galactic center. The relative number is in the logarithmic scale. Top left panel: the distribution of interaction modes of the neutron stars that interact with the ionized gas in the Minispiral region. Three basic modes are monitored: ejector (E), propeller (P), and accretor (A). The distribution alters significantly as the density increases. The temperature is fixed at $\overline{T}_{\rm{e}}=6 \times 10^3\,\rm{K}$. Top right panel: the distribution of interaction modes for the fixed density of $\overline{n}_{\rm{e}}=5 \times 10^4\,\rm{cm^{-3}}$. We vary the temperature according to the description in the plot. Only two modes, ejector and propeller, are present. The relative number of ejector modes increases $(0.9187, 0.9189, 0.9191, 0.9192, 0.9192, 0.9196)$ with increasing temperature. Bottom left panel: distribution of interaction modes of neutron stars passing through the Minispiral for four different distributions: (1) Gaussian, (2) broader Gaussian, (3) uniform distribution, and (4) combined distribution: Gaussian $+$ uniform. Parameters of individual distributions are summarized in Table~\ref{tab_distparam}. Bottom right panel: distribution of interaction modes of neutron stars for an increasing distance from the Galactic center. The magnetic field and period of neutron stars are distributed according to the combined distribution (4) in Table~\ref{tab_distparam}.}
\label{fig_histograms}
\end{figure*}

Neutron stars are distributed uniformly in the space of orbital elements, having a uniform distribution of the logarithm of the semi-major axis, $(0.05, 1.00)\,\rm{pc}$, eccentricity, $(0.0,1.0)$, and the cosine of inclination, $(-1,1)$. The number of interactions with the Minispiral arms per unit of time is $\sim 1\%$ -- $10\%$, depending on the length-scale of the individual clumps, $1$--$2$ arcsec, respectively. Thus, for an expected size of the population $\sim 10^4$, about $100$ to $1000$ members may interact with the ionized gas of the Minispiral.

Furthermore, the occurrence of interaction modes (E, P, A) is investigated according to the classification in Table~\ref{tab_interaction_modes}. We summarize the results via a series of histograms in Figure~\ref{fig_histograms} (Top left panel). It is obvious that for typical densities in the Sgr~A West region inferred from observations \citep[$10^4$--$10^5\,\rm{cm^{-3}}$, refs.][]{2010ApJ...723.1097Z,2012A&A...538A.127K} ejector mode is predominant $(\gtrsim 90\%)$; for larger densities, the fraction of propeller neutron stars increases, and at $\sim 10^7\,\rm{cm^{-3}}$ it is approximately the same as the fraction of ejectors and starts to prevail for higher densities, at which the accretor regime also becomes more prominent. This behaviour can be explained by the dependence of the stopping radius, eqs.~\eqref{eq_alfven_radius} and \eqref{eq_shvartsman_radius}, on the accretion rate: for higher densities, $\dot{M}_{\rm{c}}$ increases and the stopping radius decreases. Thus, we naturally obtain a larger fraction of propellers and accretors for higher densities. Such densities may hypothetically be found in cold, obscured regions of the Minispiral \citep{2012JPhCS.372a2019M}, in the Circumnuclear disk \citep{2005bhcm.book.....E,2010RvMP...82.3121G}, and in dense gaseous-dusty tori and disks of AGN.       

Second, we perform the simulation run with fixed density $\overline{n}_{\rm{e}}=5\times 10^4\,\rm{cm^{-3}}$. The parameters related to the magnetic field and the period distributions are kept as before. The electron temperature is varied from $1000\,\rm{K}$ to $60\,000\,\rm{K}$. We just make a note that the motivation for these density and temperature variations comes from the fact that the Minispiral consists of several streams and clumps of different properties that coexist \citep{2012A&A...538A.127K}. After several runs, we find that the distribution of interaction regimes does not vary significantly: ejectors constitute $\gtrsim 92\%$ and propellers $\lesssim 8\%$ of those neutron stars that collide with the streams, see Figure~\ref{fig_histograms} (Top right panel). For higher temperatures, the accretion rate $\dot{M}_{\rm{c}}$ drops and the stopping radius increases, see eqs~\eqref{eq_alfven_radius} and \eqref{eq_shvartsman_radius}. Hence, the number of propellers decreases whereas ejectors are more abundant. However, dependence of the distribution on temperature is rather weak. No accretors appear, since we keep the density at $5\times 10^4\,\rm{cm^{-3}}$ (compare also with Figure~\ref{fig_histograms} -- Top left panel). 

Moreover, we study the distribution of the interaction regimes of neutron stars for stratified Minispiral clumps and different distributions. To this end, we implement clumps with a variable density profile. They consist of four concentric shells with the corresponding outer radii $r_{N}=\left(\frac{1}{2}\right)^{N-1} \times 0.08\,\rm{pc}$, $N \in [1,2,3,4]$. The number density increases towards the center, starting with $10^4\,\rm{cm^{-3}}$ up to $10^7\,\rm{cm^{-3}}$, increasing inwards by one order of magnitude for each concentric shell. The temperature is set at the mean value of $\overline{T}_{\rm{e}}=6\,000\,\rm{K}$. This scheme ensures that most of the gaseous medium has number densities in the range $(10^4,10^5)\,\rm{cm^{-3}}$, which is in agreement with radio and infrared observations.

\begin{table*}[tbh!]
\centering
\begin{tabular}{ccc}
\hline
\hline
Number & Distribution type & Parameters\\
\hline
\multirow{2}{*}{$1$} & \multirow{2}{*}{Gaussian} & $\mu_{\rm{P}}=-0.2188$, $\sigma_{\rm{P}}=0.3488$\\
                     &          & $\mu_{\rm{B}}=12.0900$, $\sigma_{\rm{B}}=0.4711$\\
\hline                     
\multirow{2}{*}{$2$}  & \multirow{2}{*}{Gaussian} & $\mu_{\rm{P}}=-0.2178$, $\sigma_{\rm{P}}=0.7019$\\
                      &           & $\mu_{\rm{B}}=12.0888$, $\sigma_{{B}}=0.85098$\\
\hline                      
\multirow{2}{*}{$3$}  & \multirow{2}{*}{Uniform}  & $\log P_{\rm{min}}=-3$, $\log P_{\rm{max}}=2$\\
                      &          &  $\log B_{\rm{min}}=7$, $\log B_{\rm{max}}=15$ \\ 
\hline                      
\multirow{4}{*}{$4$}  & \multirow{4}{*}{Combined} & $\mu_{\rm{P}}=-0.2189$, $\sigma_{\rm{P}}=0.3478$\\
                      &          & $\mu_{\rm{B}}=12.09$, $\sigma_{\rm{B}}=0.47$\\ 
                      &          & $\log P_{\rm{min}}=-3$, $\log P_{\rm{max}}=2$\\
                      &          &  $\log B_{\rm{min}}=7$, $\log B_{\rm{max}}=15$ \\                      
\hline
\end{tabular}
\caption{Distributions of the period and the maximum surface intensity of the dipole magnetic field of neutron stars considered in our work.}
\label{tab_distparam}
\end{table*}

We analyse the interaction modes for four different distributions whose types and parameters are summarized in Table~\ref{tab_distparam}. Distribution (1) is inferred from the Gaussian fit of the main peak of both the period and the magnetic field distributions of observed neutron stars \citep{2005AJ....129.1993M}. Distribution (2) is a broader Gaussian distribution, (3) is a uniform distribution across a large span of values: $B \in (10^7,10^{15})\,\rm{G}$ and $P \in (10^{-3},10^2)\,\rm{s}$. Distribution (4) is a combined distribution of (1) and (3), where (3) uniformly fills an ellipse in the $PB$ plane that satisfies the following condition,

\begin{equation*}
\frac{(\log P-\mu_{\rm{P}})^2}{s_{\rm{P}}^2}+\frac{(\log B_{0}-\mu_{\rm{B}})^2}{s_{\rm{B}}^2} \leq 1\,,
\end{equation*}    
where $s_{\rm{P}}$ and $s_{\rm{B}}$ are the corresponding semi-major axes, $s_{\rm{P}}=(\log P_{\rm{max}} - \log P_{\rm{min}})/2$ and $s_{\rm{B}}=(\log B_{\rm{max}} - \log B_{\rm{min}})/2$.      

We perform Monte Carlo simulations with the distribution of the orbital elements of neutron stars as described before. The results of the study of the modes of neutron stars that interact with the Minispiral are presented in Figure~\ref{fig_histograms} (Bottom left panel). In the case of distribution (1), the ejector mode dominates, whereas the propeller and the accretor modes are represented by small amounts, $(E;P;A) \approx (95.4; 4.6; 2\times 10^{-3})\%$. For the broader Gaussian distribution (2), the ejector mode decreases, whereas the propeller and the accretor modes increase in comparison with the distribution (1), $(E;P;A) \approx (81.9; 17.8; 0.3)\%$. This trend continues in the case of the uniform distribution (3), for which the accretor mode rises considerably, $(E;P;A) \approx (60.2; 20.3; 19.5)\%$. Finally, the combined distribution (4) with a Gaussian peak results in the dominant ejector mode again, $(E;P;A) \approx (82.3; 13.4; 4.3)\%$. In all studied cases, the ejector mode dominates $(\gtrsim 60\%)$. However, the distribution of the interaction modes is very sensitive to the internal properties of neutron stars (surface magnetic field and period), which evolve with time. The purpose of the four discussed distributions of the period and magnetic field is to show general trends; more detailed studies based on the age distribution of stars in the Galactic center are required to constrain the distribution further.   

Finally, we study the change in the distribution of interaction regimes with increasing distance from the Galactic center. We use the combined distribution (4) of the period and magnetic field to analyse the general behaviour, see Figure~\ref{fig_histograms} (bottom right panel). The greater the distance, the smaller the abundance of ejectors is, whereas the number of propellers and accretors increases. This is due to the fact that the relative velocity between stars and gaseous clumps decreases for a larger distance, which implies an increase in the capture rate $\dot{M}_{\rm{c}}$, see eq.~\ref{eq_bhl2}. This is translated into shrinkage of the stopping sphere, according to eqs.~\eqref{eq_alfven_radius} and \eqref{eq_shvartsman_radius}. Hence, there is a higher probability that the gravitational capture radius and also the light radius become larger than the stopping radius, which favours the onset of the propeller and accretor stage, see Table~\ref{tab_interaction_modes}.  

\paragraph*{Distribution of interaction modes in the $PB$ plane.}      

It is useful to plot the interaction modes of neutron stars passing through the ionized Minispiral region in the $PB$ plane (period-magnetic field). We plot such a map for a uniform distribution (3; see Table~\ref{tab_distparam}) and distinguish individual regimes by colour: ejector -- red, propeller -- green, and accretor -- blue. The points in the $PB$ plane where multiple modes occur are marked by shades of these colours according to the probability of occurrence of particular regimes. 

\begin{figure}[tbh!]
\centering
\begin{tabular}{cc}
\includegraphics[width=\linewidth]{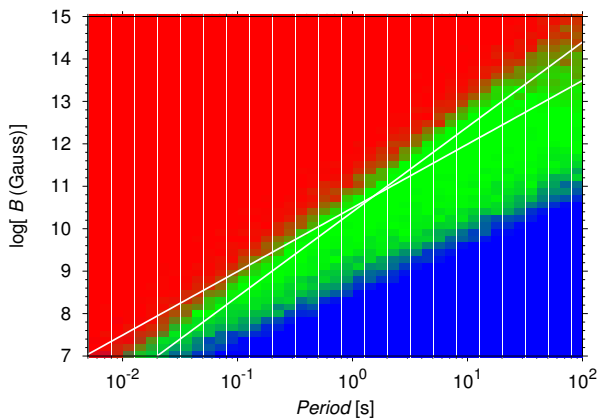} 
\end{tabular}
\caption{The uniform distribution of interacting neutron stars in the $PB$ plane. Interaction modes are color-coded: ejector--red, propeller--green, and accretor--blue. Two white lines mark the \textit{death lines}, below which pulsating neutron stars turn off.}
\label{fig_pbgraph}
\end{figure}  

The $PB$ plane is divided by a pair of \textit{death lines}, $B \approx 2.5 \times 10^{10}\, P^2\,(\rm{G})$ and $B\approx 3.1 \times 10^{10}\,P^{3/2}\,(\rm{G})$. Above these lines, electron-positron pairs can be produced in strong magnetic fields, and the pulsars are active, whereas below these lines pair production no longer proceeds and pulsars are turned off \citep{padmanabhan2000theoretical}. The ejector mode dominates the region above these lines, the accretor stage prevails below them, and the propeller mode occurs in the vicinity of these lines, see Figure~\ref{fig_pbgraph}.

\section{Discussion}
\label{sec:discussion}

Let us briefly discuss plausible observable consequences of the results presented here. 

Ejectors manifest themselves mainly as radiopulsars. The predicted number of neutron stars in the central parsec of the Galactic centre is $\sim 10^4$; the expected fraction of ejectors is $\sim 90\%$ (see Figure \ref{fig_histograms}) assuming that the dominant population of neutron stars is the same as is observed in the disk population. This fraction could be decreased if the population of old, isolated accreting neutron stars is more prominent. One should be aware of the fact that pulsars stay active for a certain amount of time, which can be approximated as $\tau \approx 100\,\rm{Myr}$, when they reach the death lines and turn off, see Fig. \ref{fig_pbgraph}. Around the death lines they become propellers, and with further spin-down and magnetic field decay they start to accrete interstellar matter. 

We take the beaming fraction of the ejectors $f_{\rm{b}}=0.2$ \citep{1989ApJ...345..931E}, i.e. pulsars that are expected to beam towards us, so potentially $\sim 1800$ pulsars could be detectable with sufficient sensitivity. However, there is an apparent lack of pulsar detections in the Galactic centre region. The pulsar searches in the Galactic disk are generally affected by high background temperature, which increases the minimum detectable flux. Moreover, all observations pointed toward the Galactic centre region suffer from interstellar scattering resulting from turbulent plasma, which causes temporal broadening of the pulses to $\sim 2000\nu_{\rm{GHz}}^{-4}\,\rm{s}$ \citep{2004hpa..book.....L} ($\nu_{\rm{GHz}}$ is the observing frequency in $\rm{GHz}$). At usual observing frequencies $\sim 1\,\rm{GHz}$ it is not possible to detect even long-period pulsars, and the surveys have to increase the frequency to $\sim 10\,\rm{GHz}$. However, since pulsars have a spectral energy distribution in the power-law form $S_{\nu} \propto \nu^{\alpha}$, where $\alpha<0$, the flux decreases for higher frequencies. To sum up, the lack of observed radiopulsars in the Galactic centre may be explained by the low sensitivity of current observing facilities. A major breakthrough is expected to come with the project of \textit{Square Kilometre Array--SKA} project \citep{2004NewAR..48..993K}, which will provide much higher sensitivity than current instruments. Mapping the close vicinity of Sgr~A* will be the task of the \textit{Event Horizon Telescope--EHT} \citep{2011ApJ...727L..36F}, which may detect pulsars strongly bound to the supermassive black hole, and the data could then be used to put the theory of general relativity to very precise tests. 

\begin{figure}[!tbh]
\centering
\includegraphics[width=\linewidth]{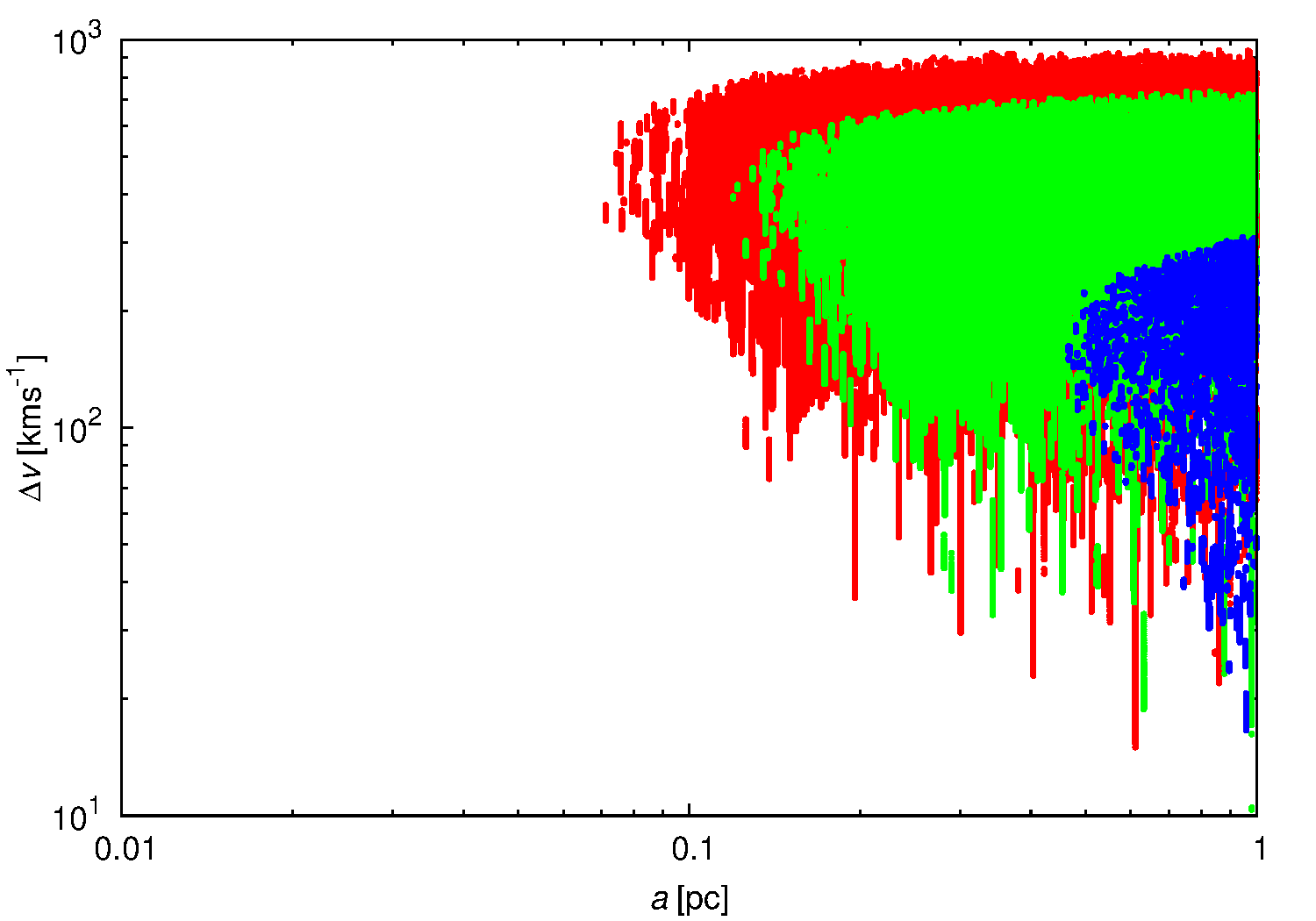}
\caption{The relative velocity of interacting neutron stars with respect to the Minispiral streams versus the semi-major axes of neutron stars. Red points label the interactions with the Northern Arm, green points stand for the Eastern Arm, and blue points represent passages through the Western Arc. }
\label{fig_relvel_ns_minispiral}
\end{figure}

Another way to detect ejectors--pulsars is to look for the cavities, pulsar wind nebulae (PWN), and bow-shock structures formed when supersonic ejecting neutron stars emitting relativistic wind propagate through the denser medium of the Minispiral. For the average electron temperature of the plasma in the Minispiral region, $\overline{T}_{\rm{e}} \approx 6 \times 10^3\,\rm{K}$ \citep{2010ApJ...723.1097Z}, the sound speed is $c_{\rm{s}} \approx (k\overline{T}_{\rm{e}}/m_{\rm{H}})^{1/2}\approx 7\,\rm{km\,s^{-1}}$. According to Fig.~\ref{fig_relvel_ns_minispiral} of the semi-major axis--relative velocity plot, typical relative velocities of neutron stars that encounter the high-density gas of the Minispiral are greater than the sound speed of the interstellar medium by one--two orders of magnitude. Hence, the motion is generally supersonic, and bow-shock structures associated with neutron stars potentially form in the Minispiral arms. The size of such a cavity is determined by the energy input, $\dot{E} \approx - \dot{E}_{\rm{rot}}= - I\Omega\dot{\Omega}$, from the central ejector, and also by the structure and the properties of the surrounding medium. The velocity of the pulsar $\mathbf{v_{\rm{rel}}}$ with respect to the ambient medium also influences the size and the orientation of the bow-shock structure. Along the relative velocity $\mathbf{v_{\rm{rel}}}$ the pressure of the wind of the ejector and the ram pressure of the ISM are in balance at the contact discontinuity. 

The distance of the contact discontinuity from the central ejector is given by the \textit{stand-off} distance \citep[e.g.,][]{2014ApJ...784..154B}:    

\begin{equation}
r_{\rm{bs}}=\left(\frac{\dot{E}}{4 \pi c \rho_{\rm{a}} v_{\rm{rel}}^2}\right)^{1/2}\,,
\label{eq_radius_pwn}
\end{equation}  
where the density of the ambient medium (the Minispiral) is $\rho_{\rm{a}} \approx m_{\rm{H}} n_{\rm{H}}$, and we adopt the values from \citet{2010ApJ...723.1097Z,2012A&A...538A.127K}; the major part of the Minispiral gaseous clumps has $n_{\rm{H}}\approx 10^4$ -- $10^5\,\rm{cm^{-3}}$, with the core of each clump of $10^7\,\rm{cm^{-3}}$. Equation \eqref{eq_radius_pwn} describes the characteristic size of both the bow-shock associated with a fast neutron star and the pulsar wind nebula, i.e. the distance of the contact discontinuity from the central pulsar. In the case of PWN, the termination shock is characterized by concentric tori and jets, as is observed in the Crab nebula, for example.

\begin{figure*}[tbh!]
\centering
\begin{tabular}{cc}
\includegraphics[width=0.5\linewidth]{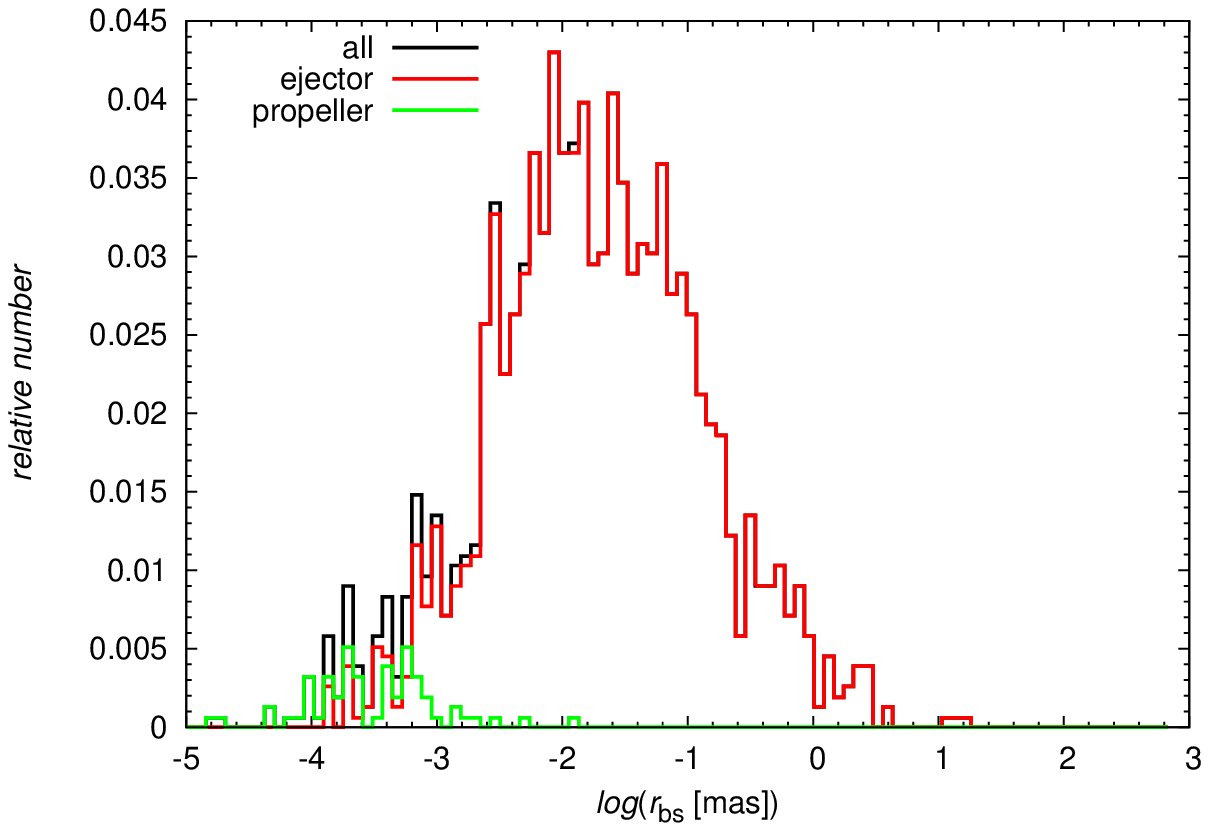} & \includegraphics[width=0.5\linewidth]{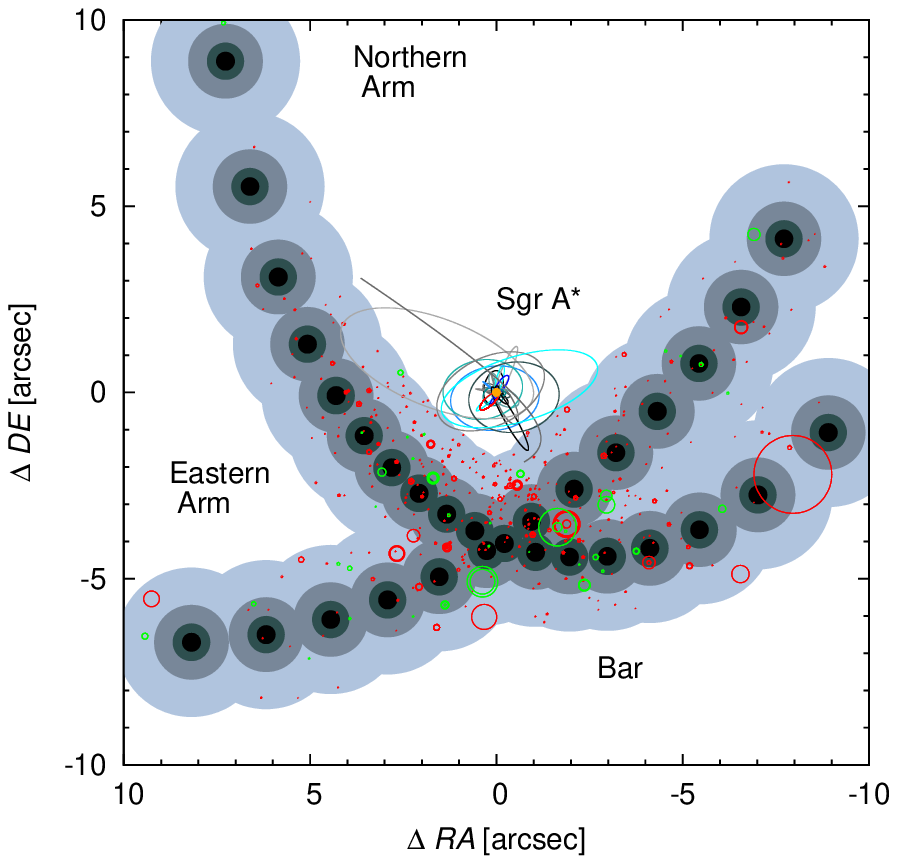}
\end{tabular}
\caption{Exemplary simulated distribution of bow-shock sizes associated with neutron stars according to Gaussian distribution (1) in Table~\ref{tab_distparam}. Left panel: histogram of bow-shock sizes. The green histogram represents propeller bow-shocks with rather small characteristic length scales that are beyond the detection limit. Right panel: comparison of bow-shock sizes in the simulated $20'' \times 20''$ image of the Minispiral. The ejector bow-shocks (red circles) are artificially enlarged by a factor of $100$ and propeller bow-shocks are enlarged $100\,000$ times enlarged for clarity. The elliptical orbits of selected S-stars are also plotted for illustration \citep{2009ApJ...692.1075G}.}
\label{fig_bowshock}
\end{figure*}

We compute the typical sizes of bow shocks and/or PWN for an ensemble of $10\,000$ neutron stars, using eq. \eqref{eq_radius_pwn} and the assumption that the Gaussian distribution (1) in Table~\ref{tab_distparam} is approximately valid for the Galactic center population. Assuming that $\gtrsim 90\%$ of interacting pulsars are ejectors, about one thousand (in our exemplary runs $\sim 1500$) bow-shock structures could be present along the Minispiral arms. However, most of them are rather small $(< 1\,\rm{mas})$, and are therefore below the detection limit. Propeller bow shocks are all below $\sim 0.01\,\rm{mas}$, whereas the bow-shock structures of ejectors have sizes above this value and up to $\sim 10\,\rm{mas}$, see Fig.~\ref{fig_bowshock} (Left panel) for comparison; the values along the horizontal axis are decadic logarithms of stand-off radii in milliarcseconds.

The number of larger bow-shock structures $(\gtrsim 1\,\rm{mas})$ is $\sim 30$ for $\sim 1500$ neutron stars that interact with the Minispiral out of $\sim 10\,000$ neutron stars that occupy the innermost parsec. However, only $\sim 10$ of them are greater than $10\,\rm{mas}$. Fig.~\ref{fig_bowshock} (right panel) compares their sizes with the whole Minispiral structure, as seen from the Earth for an exemplary simulation.

In addition, neutron stars are expected to occur closer to the SMBH, where relativistic effects could be detected. They may be members of the S-cluster, where they were originally locked in binary systems with hot stars. Close pericenter passages can, however, disrupt such binary systems due to triple scattering. The components become separated, and for a certain set of orbital parameters, one component can escape the system on a hyperbolic orbit \citep[see ][for details]{2014A&A...565A..17Z}. These hypervelocity neutron stars could potentially be seen to interact with the gaseous environment of Sgr~A West and further away.

The HII region of Sgr~A West, especially the dense Minispiral arms and the bar, seem to be a promising target to search for the effects of interaction with passing compact objects, both for the number of encounters and for the observable effects. Due to its large size, the Minispiral appears to be a more favourable target for detecting the manifestation of the neutron star population than an isolated clump, such as G2/DSO currently observed in the Galactic center \citep{2013MNRAS.435L..19D}.

\section{Conclusions}
\label{sec:conclusions}

We have studied the distribution of interaction modes of the neutron star population in the Galactic center. We have explained that the established mode of interaction depends on the intrinsic properties of a neutron star (the rotational period and the magnetic dipole surface intensity) and the external conditions (parameters of the gaseous medium: density and temperature). 

We focused on the distribution of three main interaction regimes of neutron stars: ejector (E),  propeller (P), and accretor (A). We found that the distribution of the modes is strongly dependent on the plasma density, while it is only weakly dependent on the temperature in the range $\sim (10^3,10^4)\,\rm{K}$. This further implies that the distribution varies across different galaxy types.         

We went on to propose an alternative way to look for ejecting neutron stars in the Galactic center. One can look for the signatures of the interaction of neutron stars with the dense Minispiral arms. Common structures associated with the supersonic motion of pulsars are bow-shocks. Our analysis shows that the number of relatively large structures in this region, $\gtrsim 10\,\rm{mas}$, can be of the order of $\sim 10$. However, we stress that the results depend on the assumed distribution of the rotational periods and the surface magnetic fields. The comparison of different synthetic distributions of bow-shock sizes with the observed structures will help to constrain the parameters of the Galactic center population of young neutron stars. These bow-shocks are expected to be sources of polarized non-thermal radiation, possibly detectable in the X-ray, infrared, and radio wavelengths. Several cases of such comet-shaped sources are indeed observed in this region \citep{2013ApJ...777..146Z,2010A&A...521A..13M}. Further analysis is required to strengthen or reject their association with propagating neutron stars. In our upcoming paper we will further explore the observability of these features (Zaja\v{c}ek et al., in preparation).



\begin{acknowledgements}
The authors acknowledge the support from the Grant Agency of the Charles University in Prague (grant \# 879113, ``The study of neutron stars in the galactic center") and also funding from an ongoing grant GACR - Albert Einstein Center for Gravitation and Astrophysics No. 14-37086G. Further support has been received from the DAAD-Czech joint research program ``Effects of Albert Einstein's Theory of General Relativity Revealed with the Instruments of European Southern Observatory".
\end{acknowledgements}

\bibliographystyle{actapoly-astro}

\end{document}